\documentclass[aps,prl,preprint,groupedaddress,amsmath,amssymb]{revtex4-1}
\usepackage{amsmath}
\usepackage{graphicx}
\usepackage{dcolumn}
\usepackage{bm}

\begin{document}

%\preprint{}



\title{Double-relativistic-electron-shell laser proton acceleration}

\author{Yongsheng Huang}
 \homepage{http://www.anianet.com/adward}
\email{huangyongs@gmail.com}
\affiliation{China Institute of Atomic Energy, Beijing 102413, China.}%
\author{Naiyan Wang }
\author{Xiuzhang Tang }
\author{Yijin Shi}
\affiliation{China Institute of Atomic Energy, Beijing 102413,
China.}

\author{Yan Xueqing}
\affiliation{Institute of Heavy Ion Physics, Peking University,
Beijing 100871, China}

\author{Zhang Shan}
\affiliation{Beijing Normal University, Beijing 100875, China}
%\thanks{}
%\affiliation{}


\date{\today}

\begin{abstract}
A new laser-proton acceleration structure combined by two
relativistic electron shells, a suprathermal electron shell and a
thermal electron cloud is proposed for $a\gtrapprox80\sigma_0$,
where a is the normalized laser field and $\sigma_0$ is the
normalized plasma surface density. In the new region, a uniform
energy distribution of several GeV and a monoenergetic
hundreds-of-MeV proton beam have been obtained for $a=39.5$. The
first relativistic electron shell maintains opaque for incident
laser pulse in the whole process. A monoenergetic electron beam has
been generated with energy hundreds of MeV and charge of hundreds of
pC. It is proposed a stirring solution for relativistic
laser-particle acceleration.
\end{abstract}

%\keywords{}

\maketitle

%\section{\label{sec:level1}Introduction}
Laser-ion acceleration has been an international research
focus\cite{MakoTajima,Machnisms,Esirkepov}, however it is still a
challenge to obtain mono-energetic proton beams larger than
$100\mathrm{MeV}$. Although the field in the laser-plasma
acceleration is three to four orders higher than that of the classic
accelerators, it decreases to zero quickly in several pulse
durations for target normal sheath acceleration
(TNSA)\cite{Machnisms} for $a\ll\sigma_0=\frac{n_el_e}{n_c\lambda}$,
where $a=eE_l/m\omega c$, $E_l$ is the electric field of the laser
pulse, $\omega$ is the laser frequency, $e$ is the elementary
charge, $m$ is the electron mass, $c$ is the light velocity, $n_0$
is the initial plasma density, $n_c$ is the critical density,
$\lambda$ is the wave length, $n_e$ is the electron density, $l_e$
is the thickness of the electron shell. As a promising method to
generate relativistic mono-energetic protons, radiation pressure
acceleration (RPA) has attracted more
attention\cite{Esirkepov,Henig,YanxqPRL,unlimitedRPA} and becomes
dominant in the interaction of the ultra-intense laser pulse with
thin foils if $a\approx\sigma_0$. Even in the unlimited ion
acceleration\cite{unlimitedRPA}, only the ions trapped in the
electron shell can obtain efficient acceleration, therefore, the
total charge of the ion beam is quite limited due to the transverse
expansion\cite{unlimitedRPA}. Although in RPA region, the energy
dispersion will become worse with time.

Fortunately, for $a\gtrapprox80\sigma_0$, in the relativistic case,
a new acceleration region appears: double relativistic electron
shells come into being. The ions between the two electron shells
will be accelerated most efficiently and obtain a uniform energy
distribution. The first electron shell is ultra-relativistic and is
totally separated from the ions. The second electron shell comes
into being in the potential well induced by the electron
recirculation and will also be relativistic. It is in the ion region
and follows the ion front and forms a potential well which traps
energetic ions and accelerates them to be quasi-mono-energetic and
relativistic. Following the second electron shell, a suprathermal
electron beam comes into being and induces another potential well
which can also trap lots of ions and accelerate them to obtain a
monoenergetic relativistic one. On the whole, the maximum ion energy
can reach several GeV and a relativistic monoenergetic ion beam with
relative energy dispersion smaller then $5\%$ can be obtained.

As a ultraintense laser pulse is shot on a ultra-thin plasma foil
for $a\gtrapprox80\sigma_0$, the electron shell is compressed to
ultra-high density and pushed forward to be separated from the ion
shell totally, and gains ultra-relativistic energy that can make
sure it opaque for the laser pulse. In the whole process, the first
electron shell keeps opaque for the incident laser pulse and is
pushed by it continuously.  According to Eq.
(22)\cite{unlimitedRPA}, it can be satisfied that the opaqueness
condition of electron shell for laser in the acceleration:
\begin{equation}\label{eq:opaque}
a_0\leq \pi(\gamma_e+p_e)\hat{n}_e\hat{l}_e,
\end{equation}
since $d\ln (p\hat{n}_e\hat{l}_e)/dt>0$, as pointed by Bulanov,
$p\propto t^{1/3}$ is the normalized electron momentum, and
$n_el_e=n_0l_0$ in the no transverse expansion case, where the
electron density $n_e, n_0$ are normalized by $n_c$ and $l_e$ is
normalized by $\lambda$, $a_0=eE_0/m\omega_0 c$, $E_0$ is the
electric field of the laser pulse, $\omega_0$ is the laser
frequency, $e$ is the elementary charge, $m$ is the electron mass,
$c$ is the light velocity, $n_0$ is the initial plasma density,
$n_c$ is the critical density, $\lambda$ is the wave length. In the
ultra-relativistic case, the r.h.s. of Eq. (\ref{eq:opaque}) is
approximate $2\pi p\hat{n}_e\hat{l}_e$.

\begin{figure}
{
\includegraphics[width=1\textwidth]{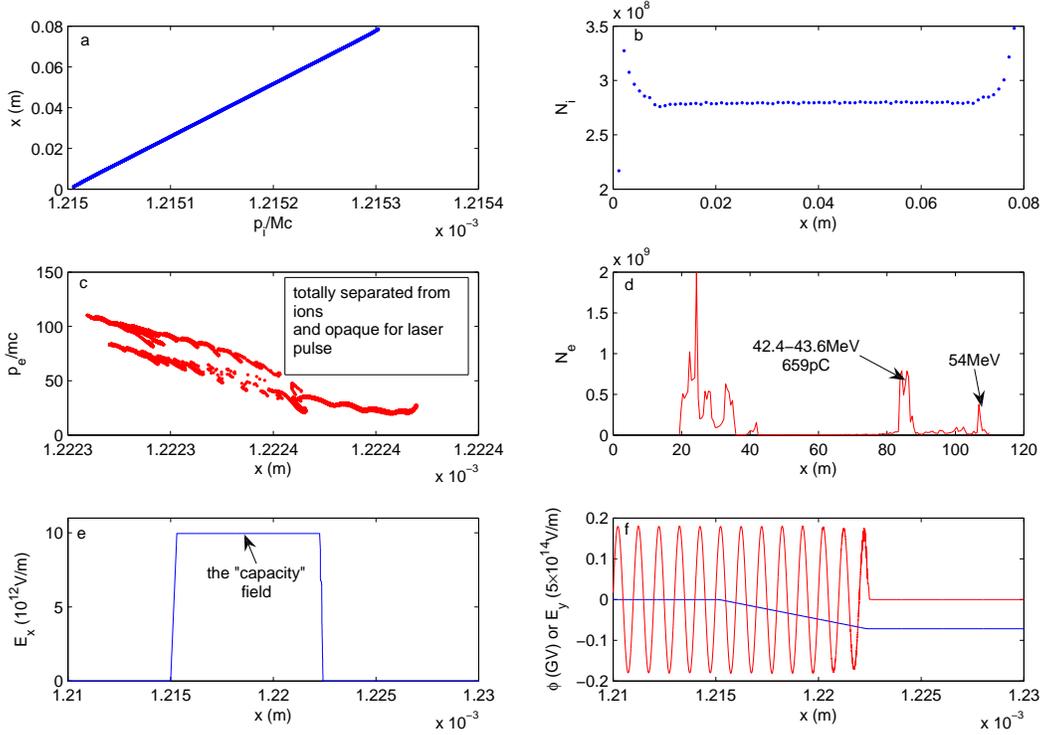}% Here is how to import EPS art
}  \caption{\label{fig:PIC39-25} (Color online) Simulation results
by one-dimensional VORPAL at $t=25$fs: the first opaque relativistic
electron shell forms and is totally separated from ions. (a) and
(c): the phase space distribution of ions and electrons. The
electron shell is totally separated from the ions and is
relativistic and opaque for laser pulse. (b) and (d): the density
distribution of the ions and electrons with normalized momentum. The
ion energy distribution is uniform. The electron energy distribution
contains several monoenergetic ones. (e): the longitudinal space
charge separation field, which is similar to the field in a
capacity, is uniform and is $9.95\times10^{12}$V/m. (f): the
potential and laser field. The electron shell is relativistic and
opaque for the laser pulse. }
\end{figure}

For $a=39.5$ and $\hat{n}_e\hat{l}_e=49\times0.01=0.49$,
$a\approx81\hat{n}_e\hat{l}_e$. When the laser pulse interacts with
the plasma shell, the electrons are compressed to high density and
pushed forward to be separated from the ion shell. As shown by
Figure \ref{fig:PIC39-25} (c) and (d), at $t=25\mathrm{fs}$, the
normalized momentum of the electron shell reaches $20-100$. With Eq.
(\ref{eq:opaque}), the compressed high-density electron shell is
opaque for the laser pulse as shown by Figure \ref{fig:PIC39-25}
(f). The electrons will be accelerated efficiently and continuously
by the radiation pressure of the laser.

This high-density relativistic electron shell is called the first
relativistic electron shell. Between the first electron shell and
the ions, a uniform space-charge separation field forms and
accelerates the ions at the ion front and drags the electrons as
shown by Figure \ref{fig:PIC39-25} (e). The "capacity" field is
decided by the surface density of the electron shell:
\begin{equation}\label{eq:Ecap}
E_{cap}=\frac{en_el_e}{\epsilon_0},
\end{equation}
For $n_e=5.5\times10^{22}\mathrm{/cm^3}$, $l_e=10\mathrm{nm}$, the
stable field is $9.95\times10^{12}\mathrm{V/m}$, which accelerates
the ions at the rear of the ion shell continuously. Therefore the
maximum ion energy is proportional to the acceleration length,
$d_{acc}$,
\begin{equation}\label{eq:Eiond}
E_i=E_{cap}d_{acc}(\mathrm{eV}),
\end{equation}
before the electron shell breaks up.

\begin{figure}
{
\includegraphics[width=1\textwidth]{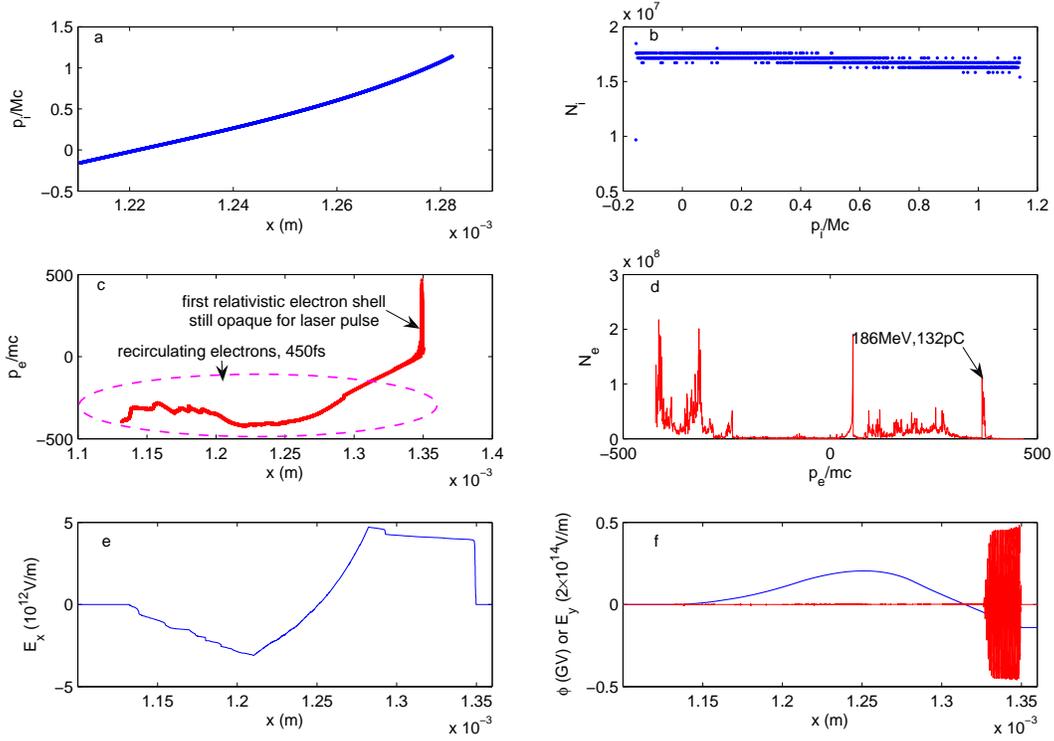}% Here is how to import EPS art
}  \caption{\label{fig:PIC39-500} (Color online) Simulation results
by one-dimensional VORPAL at $t=450$fs: electron recirculation
begins and generates a potential well for electrons, where $p_i,
p_e$ is momentum of ion and electron respectively, $M$ is the proton
mass. (a) and (c): the phase space of ions and electrons
respectively. The electron recirculation happens and some of them
move in the opposite direction relativistically. (b) and (d): the
energy distribution of ions and electrons respectively. The first
electron shell is accelerated most efficiently and continuously. (e)
the longitudinal field decreases due to the electron recirculation.
However it is still uniform between the ion front and the first
electron shell. (f): a potential hill for ions and a potential well
for electrons forms due to the electron recirculation. The first
relativistic electron shell maintains opaque for the laser pulse.}
\end{figure}

After several hundreds of femtoseconds, some electrons leak out from
the electron shell continuously and move backward and round again
and follow the ion front, however, they can not catch the ion front.
Figure \ref{fig:PIC39-500} (c), (e) and (f) shows the electron
recirculation, the decrease of the separation field, and the
formation of the potential well for electrons at $t=450\mathrm{fs}$
respectively. The normalized maximum electron momentum reaches about
$500$ as shown by Figure \ref{fig:PIC39-500} (d). Figure
\ref{fig:PIC39-500} (f) shows that the deepness of the potential
well for electrons increases about to $0.3\mathrm{GeV}$. From Figure
\ref{fig:PIC39-500} (a) and (b), the maximum ion energy is about
$500\mathrm{MeV}$. The acceleration length is about $65\mathrm{\mu
m}$. A monoenergetic electron beam of $186\mathrm{MeV}$ and $132$pC
is obtained as shown by Figure \ref{fig:PIC39-500} (d). It is
obvious that the first relativistic electron shell is still opaque
for the laser pulse.

\begin{figure}
{
\includegraphics[width=1\textwidth]{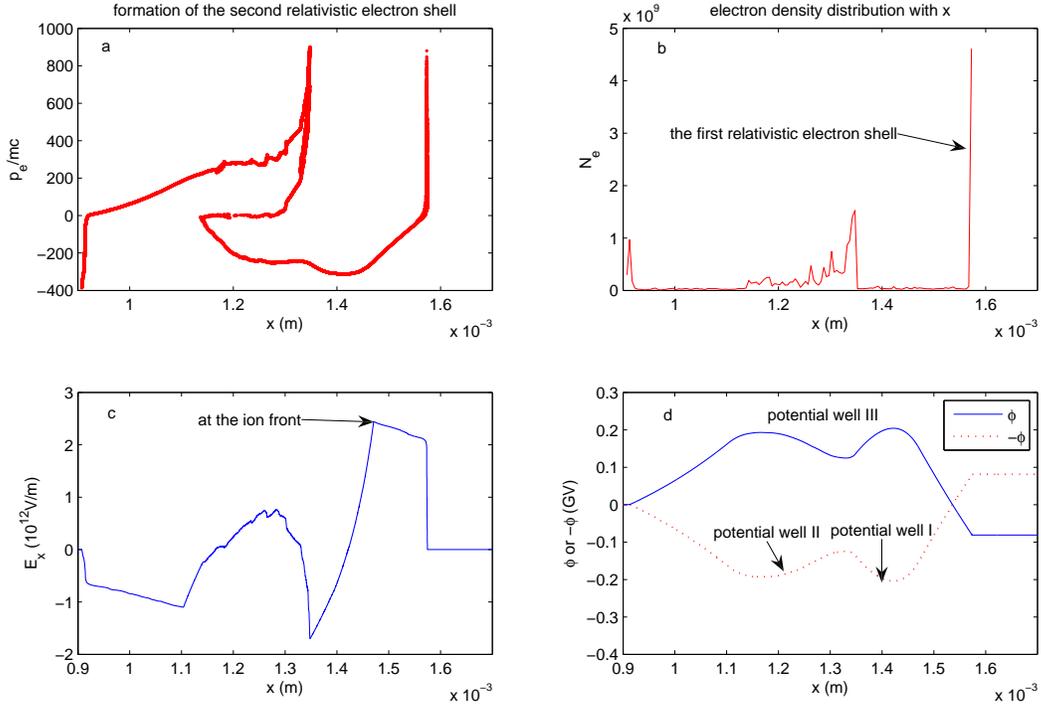}% Here is how to import EPS art
}  \caption{\label{fig:PIC39-1250} (Color online) Simulation results
by one-dimensional VORPAL at $t=1.25$ps: the second relativistic
electron shell forms and traps ions to be accelerated to
relativistic. (a): the second relativistic electron shell forms in
the potential well. It drives up the potential and forms two
potential well I and II for electrons in front of and behind itself,
a potential well III for ions at the local position of itself. It is
shown clearly in (d). (b): the number density distribution of
electrons. (c): the electron recirculation decreases the
longitudinal field continuously. It is still uniform between the ion
front and the first electron shell.}
\end{figure}

\begin{figure}
{
\includegraphics[width=1\textwidth]{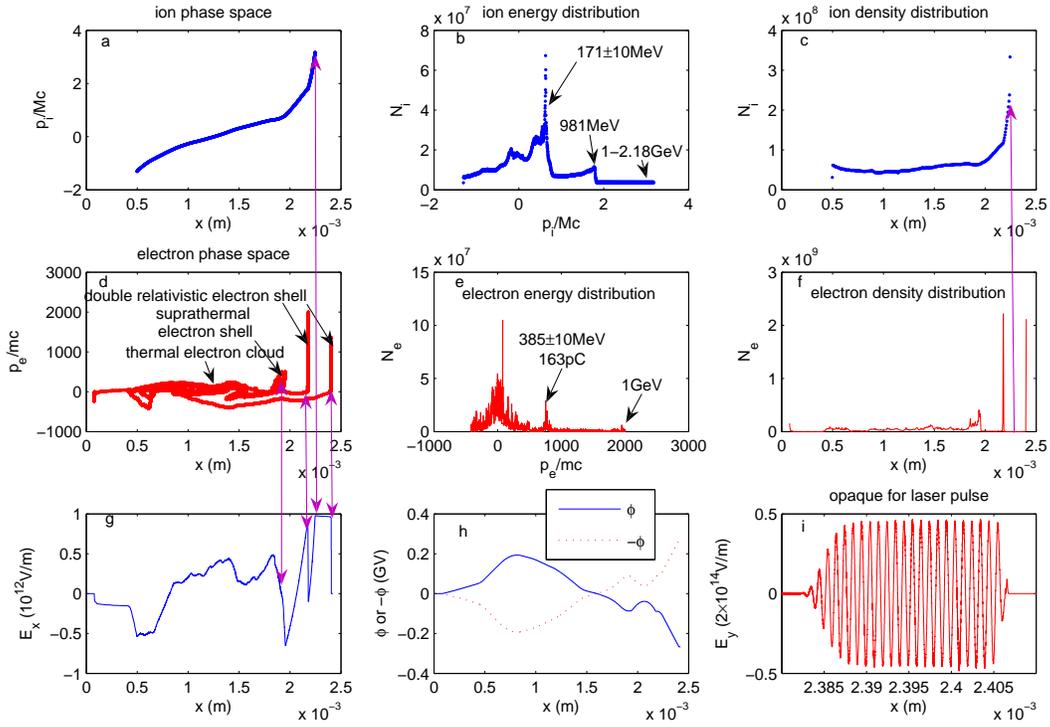}% Here is how to import EPS art
}  \caption{\label{fig:PIC39-4025} (Color online) Simulation results
by one-dimensional VORPAL at $t=4.025$ps: the thermal electron cloud
and the suprathermal electron shell come into being. (a) and (d):
the phase-space of ions and electrons respectively. The electrons
contain four main parts: the double relativistic electron shells,
the suprathermal electron shell, the thermal electron cloud. (b) and
(e): the energy distribution of ions and electrons respectively. A
monoenergetic ion beam with energy of $171\pm10$MeV is obtained by
the suprathermal electron shell. Trapped and accelerated by the
second relativistic electron shell, the ion energy distribution
drops down at $981$MeV. (c) and (f) the number density of ions and
electrons respectively. (g) the longitudinal field. (h) the
potential for ions and electrons. (i) the laser pulse field. The
first electron  shell maintains opaque for laser pulse. }
\end{figure}

As shown by Figure \ref{fig:PIC39-1250} (d), the recirculating
electrons cumulate and drive up the electron potential. In front of
and behind the accumulating electrons, two potential wells are
forming for electrons.  The potential well I still traps and
accelerates the accumulating electrons to generate the second
relativistic high-density electron shell. In the potential well II,
some electrons at the end of the second relativistic electron shell
drop into it and will be trapped and accelerated to form a
suprathermal electron shell as shown by Figure \ref{fig:PIC39-4025}
(d). At the same time and at the local position of the second
relativistic electron shell, potential well III for ions traps lots
of ions and accelerates them to be relativistic and monoenergetic.

\begin{figure}
{
\includegraphics[width=1\textwidth]{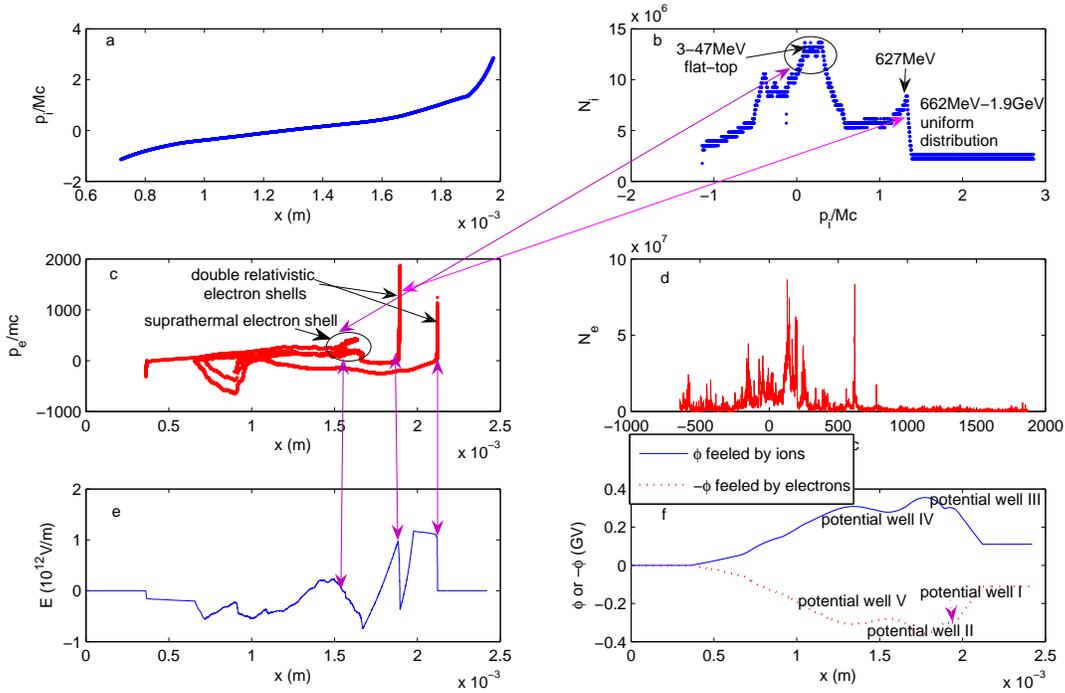}% Here is how to import EPS art
}  \caption{\label{fig:PIC39-3075} (Color online) Simulation results
by one-dimensional VORPAL at $t=3.075$ps: the suprathermal electron
shell forms and traps ions to obtain a flap-top $3-47$MeV energy
distribution. (a) and (c): the phase space of ions and electrons.
The suprathermal electron shell forms in the potential well II and
then potential V for electrons and potential IV are generated. (b)
and (d) the energy distribution of ions and electrons. A monoenergy
ion beam with energy of $627$MeV is obtained in the second
relativistic electron shell. (e): the longitudinal field induced by
the double electron shell. (f): the potential IV for ions is induced
by the suprathermal electron shell and traps ions and improves the
energy dispersion of the $3-47$MeV ion beam. The potential V can
trap slow electrons and thermalize them to obtain thermal electron
cloud. Potential III is nearly filled to be flat and the energy
dispersion of the 627MeV monoenergetic ion beam will become worse.}
\end{figure}

With time, the slow recirculating electrons can also be trapped in
potential well V as shown in Figure \ref{fig:PIC39-3075}(f) and the
third suprathermal electron shell forms in the potential well II as
shown by Figure \ref{fig:PIC39-4025}. At the position of the shell,
potential well IV traps the ions and accelerates them to obtain a
quasi-monoenergetic distribution of $171\pm10\mathrm{MeV}$. Behind
the shell, a potential well for electrons traps them and a thermal
electron cloud is generated. The ions between the double
relativistic electron shell have a uniform distribution from
$1\mathrm{GeV}$ to $2.18\mathrm{GeV}$. Trapped by the second
relativistic electron shell, the maximum energy reach
$981\mathrm{MeV}$. As shown in Figure \ref{fig:PIC39-3075} (f), the
potential well III has been filled up nearly, then the energy
dispersion will become worse. The ions with larger energy will coast
down the following potential slope and get into the ion beam between
the double relativistic electron shell. The ion number has a steep
descent for the energy larger than $981\mathrm{MeV}$ and has a slow
drop for the energy smaller than $981\mathrm{MeV}$. In the electron
energy distribution, there is a monoenergetic one of
$385\pm10\mathrm{MeV}$ and $163\mathrm{pC}$, a ultra-relativistic
one of $1\mathrm{GeV}$ and a Maxwellian one which contains the
thermal electron cloud and the suprathermal electron shell. As shown
by Figure \ref{fig:PIC39-4025}(c) and (f), the ion front is between
the double electron shell. The ions between the double electron
shell coast down the potential slope and obtain relativistic energy
as shown in Figure \ref{fig:PIC39-4025}(h) and (b).

In conclusion, the double relativistic electron shells, the
suprathermal electron shell and the thermal electron cloud induce a
new region of laser particle acceleration. In the process, several
potential wells for ions and electrons are generated. On the whole,
the double relativistic electron shells induce two relativistic
platforms of the ion energy distribution. The suprathermal electron
shell traps and accelerates a monoenergetic ion beam with several
hundreds of MeV, whose relative energy dispersion is near $5\%$.
Together with the thermal electron cloud, a thermal Maxwellian ion
beam has been obtained.

\begin{acknowledgments}
The authors  would like to thank Dr. Hong-Yu Wang for useful
discussion. The computation was carried out at the HSCC of Beijing
Normal University. This work was supported by the Key Project of
Chinese National Programs for Fundamental Research (973 Program)
under contract No. $2011CB808104$ and the Chinese National Natural
Science Foundation under contract No. $10834008$.
\end{acknowledgments}

\end{document}